\documentclass[preprint,prb,showpacs,preprintnumbers,amsmath,amssymb]{revtex4}

\usepackage{graphicx}
\usepackage{dcolumn}
\usepackage{bm}

\newcommand{\vol}[1]{\mathrm{d}#1}

\begin{document}


\title{Soft effective interactions between weakly charged 
       polyelectrolyte chains}

\author{M.\ Konieczny}
  \email{kon@thphy.uni-duesseldorf.de}
\author{C.\ N.\ Likos}
\author{H.\ L\"owen}
\affiliation{
Institut f\"ur Theoretische Physik II, 
Heinrich-Heine-Universit\"at D\"usseldorf, 
Universit\"atsstra{\ss}e 1, 
D-40225 D\"usseldorf, Germany}

\date{{\bf \today}, submitted to {\sl The Journal of
Chemical Physics}}

\begin{abstract}

We apply extensive Molecular Dynamics simulations and analytical
considerations in order to study the conformations and the effective
interactions between weakly charged, flexible polyelectrolyte chains
in salt-free conditions. We focus on charging fractions lying below
20\%, for which case there is no Manning condensation of counterions
and the latter can be thus partitioned in two states: those that are
trapped within the region of the flexible chain and the ones that are
free in the solution. We examine the partition of counterions in
these two states, the chain sizes and the monomer distributions
for various chain lengths, finding that the monomer density follows
a Gaussian shape. 
We calculate the effective interaction between the centers of mass of 
two interacting chains, under the assumption that the chains can be 
modeled as two overlapping Gaussian charge profiles.
The analytical calculations are compared with measurements from
Molecular Dynamics simulations. 
Good quantitative 
agreement is found for charging fractions below 10\%, where the chains 
assume coil-like configurations, whereas deviations develop for charge 
fraction of 20\%, in which case a conformational transition of the 
chain towards a rodlike configuration starts to take place.

\end{abstract}



\maketitle

\section{
\label{sec:intro}
Introduction
}

Polyelectrolytes (PE's) are polymer chains that carry ionizable groups
along their backbone.\cite{oosawa, dautzenberg, barrat:95}
Thereby, when dissolved in a polar solvent
such as water, these groups dissociate and a chain that carries
charges along its monomer sequence results, leaving behind the
dissociated counterions in the solution. Upon addition of salt,
additional counterions as well as coions can be found in the system.
One deals therefore with a statistical-mechanical problem that entails
aspects pertinent to both charge-stabilized colloidal 
suspensions\cite{lowen:review} and polymer solutions. This dual
character of polyelectrolytes lies in the heart of the 
difficulties in devising suitable theoretical approaches to tackle
questions regarding their conformations and structure in 
nonvanishing concentrations.
Indeed, the long-range character of the Coulomb interaction renders
the typical renormalization-group techniques that have proven very
fruitful for neutral polymer chains rather inadequate for the
problem at hand. Moreover, the appearance of additional length scales
(such as the Bjerrum length, to be defined in what follows) has
the consequence that several microscopic parameters have an influence
on the physics of the system, in contrast to the case of neutral
polymer chains.\cite{barrat:95} Simulation approaches, pioneered
by the work of Stevens and Kremer,\cite{stevens:prl:93,stevens.jcp103}
have thus proven very useful in shedding light into the question
of the typical sizes and conformations of isolated PE chains.

Considerable work has been devoted to the study of sizes, conformations
and charge distribution of isolated PE chains under the influence of
various physical parameters. These include the effects of counterions
and salt,\cite{vasilevskaya:00} of the temperature,\cite{kantor:prl:99}
of annealed versus quenched backbone charges\cite{castelnovo} as
well as of solvent 
quality.\cite{yethiraj:shew:jcp:99, micka:langmuir:99, essafi:jp:95}
In the absence of screening, it has been found that the typical size
$R$ of a PE chain scales with the degree of polymerization $N$ as
$R \sim N(\ln N)^{1/3}$ at room temperature.\cite{migliorini, liao}
Another issue concerning isolated PE chains that has attracted considerable
interest, is that of Manning condensation\cite{manning:jcp:69} along
the stiff, rodlike backbone. The fraction of condensed 
counterions has been examined both in
theory\cite{liu:macrom:99, deserno:macrom:00} and in 
experiments.\cite{rehahn:jpcb:00, rehahn:epje:01}
Counterion
condensation and fluctuations have been shown to lead
to a collapse of the rodlike chains, due to the effective monomer-monomer
attractions that result from the condensed counterions that position
themselves between the charged 
monomers.\cite{winkler:prl:98, klein:prl:98, schiessel:macrom:98, schiessel:macrom:99, golestanian:prl:99, iff33}

Relatively less is known about the behavior of polyelectrolyte 
solutions at finite concentration. Theoretical approaches on this
topic are usually based on integral equation theories that assume
rodlike configurations of the chains\cite{yethiraj:shew:prl:96, yethiraj:shew:jcp:97, yethiraj:shew:jcp:98, yethiraj:shew:jcp3:99, harnau:jcp:00} and
make use of the polymer reference interaction site model (PRISM)
of Schweizer and Curro.\cite{prism} An approach that combines aspects 
from field-theoretical methods and liquid-state theory has been developed 
by Yethiraj,\cite{yethiraj:prl:97} and which allows for a self-consistent
calculation of the solvation potential on a chain, due to the rest
of the solution.\cite{yethiraj:jcp:98} This theory has been applied
to calculate the extent of the PE chains as well as the
correlation functions between the monomers at varying concentration,
both ignoring\cite{yethiraj:prl:97, yethiraj:jcp:98} and 
including\cite{yethiraj:shew:jcp2:99, yethiraj:shew:jcp:00} 
the excluded-volume interactions. Recently, phase separation and
Coulomb criticality in polyelectrolyte solutions was also studied
by means of Monte Carlo computer simulations.\cite{thanassis:prl:03}

A very useful theoretical tool that greatly facilitates the
analysis of the structure and thermodynamics of complex fluids
is that of the effective interaction between suitably chosen 
degrees of freedom that characterize the macromolecule as a 
whole.\cite{cnl.habil} For instance, it has been recently shown
through extensive computer simulations
that neutral polymer chains can be modeled as ultrasoft colloids
if their centers of mass are chosen as the effective 
coordinates that describe the chains in a coarse-grained
fashion.\cite{louis:prl, louis:pre, louis:jcp} Effective
interactions between polyelectrolytes have not been derived
up to now, however, with the exception of the case of 
rodlike molecules.\cite{carri:jcp:99} It is the purpose of
this paper to fill this gap, by considering 
{\it weakly charged} PE chains for which (a) the chains
are still flexible and assume a large number of conformations
and (b) there are no Manning-condensed counterions. We perform
this coarse-graining procedure by tracing out the monomer
and counterion degrees of freedom in a salt-free solution
and we derive an effective, density-dependent potential of
mean force between the centers of mass of the PE chains.
Since the analytic considerations for the derivation of the
effective interaction are based on knowledge of the sizes
and shapes of individual chains, we present first in
Sec.\ \ref{sec:iso} simulation and analytical results pertaining
to single polyelectrolytes and demonstrate the good agreement
between the two. The theory and the simulation results for
the effective potential are presented in Sec.\ \ref{sec:int}, whereas
in Sec.\ \ref{sec:con} we summarize and conclude. Some technical 
derivations are shown in the Appendix.

\section{
\label{sec:iso}
Sizes and conformations of isolated PE chains
}

\subsection{
\label{sec:iso.md}
Simulation model
}

We start with the description of our simulation model, valid for both an 
isolated PE chain and two interacting PE chains. We performed monomer-resolved 
molecular dynamics simulations (MD). We employed a cubic simulation box with 
periodic boundary conditions and typical edge lengths in the range 
$L=2R_{\rm e}$ to $5R_{\rm e}$, where $R_{\rm e}$ is the end-to-end
distance of the chain, depending on the other parameters.
To integrate the equations of motion, we used the so-called velocity form of 
Verlet's algorithm.\cite{allen,frenkel,rapaport} In order to stabilize the 
temperature of the system, a Langevin thermostat was 
applied.\cite{kremer.jcp92,white.mp99,messina.jcp117} 
This method is based on the introduction of convenient random and friction 
forces taking the fluctuation-dissipation theorem into account. The electrostatic 
Coulomb interaction (see below) of charged particles was treated using Lekner's 
summation method,\cite{lekner} where an enhancement of the convergence properties 
of the associated sums is achieved via their adequate re-writing. Thus, a cut-off 
becomes feasible. To limit the memory consumption of the tables containing 
the corresponding forces, we computed the latter by means of a trilinear 
interpolation,\cite{num.recip} while simultaneously decreasing the number of grid 
points. 

The PE chains were modeled as bead spring chains of $N$ Lennard-Jones (LJ) particles. 
This method was first introduced in investigations of neutral polymer chains 
and stars\cite{stevens.jcp103,grest.mm20,grest.mm27} and turned out to be a 
reasonable approach. To mimic good solvent conditions, we use a shifted LJ 
potential to describe the purely repulsive excluded volume interaction of the 
monomers:
\begin{equation}
\label{eq:vlj}
V_{\rm LJ}(r)=
\begin{cases}
4\varepsilon_{\rm LJ}\left[\left(\frac{\sigma_{\rm LJ}}{r}\right)^{12}
-\left(\frac{\sigma_{\rm LJ}}{r}\right)^{6}
+\frac{1}{4}\right],&r\leq 2^{1/6}\sigma_{\rm LJ}\\
0,&r>2^{1/6}\sigma_{\rm LJ}.
\end{cases}
\end{equation}
Here, $r$ is the spatial distance of two interacting particles. $\sigma_{\rm LJ}$ 
denotes their microscopic length scale, and $\varepsilon_{\rm LJ}$ sets the energy 
scale for the system. We chose the value $T=\varepsilon_{\rm LJ}/k_{\rm B}$ for the 
system's temperature, where $k_{\rm B}$ is the Boltzmann constant. 

The bonds of adjacent monomers were depicted via a finite extension nonlinear 
elastic (FENE) potential:\cite{grest.mm20, grest.mm27}
\begin{equation}
V_{\rm FENE}(r)=
\begin{cases}
-\frac{k_{\rm FENE}}{2}\left(\frac{R_0}{\sigma_{\rm LJ}}\right)^2\ln
\left[1-\left(\frac{r}{R_0}\right)^2\right],&r\leq R_0\\
\infty,&r>R_0,
\end{cases}
\end{equation}
where $k_{\rm FENE}$ is the spring constant. Its value was set to 
$k_{\rm FENE}=7.0\varepsilon_{\rm LJ}$. The FENE interaction diverges at $r=R_0$, 
which determines the maximum relative displacement of two neighboring beads. 
The energy $\varepsilon_{\rm LJ}$ is the same as in Eq.\ \eqref{eq:vlj}, whereas
for the length scale $R_0$ we have chosen the value $R_0=2.0\sigma_{\rm LJ}$.

Each chain is charged by a fraction $\alpha$ in a periodical manner: every 
$1/\alpha$ bead carries a monovalent charge. Due to the requirement of 
electroneutrality, the same amount of also monovalent but oppositely charge ions, 
namely $N_{\rm c}=\alpha N$ counterions, are included in the simulation
box. They are able to freely move in the box, thereby they had to be simulated 
explicitly.

Finally, the Coulomb interaction $V_{\rm Coul}(r)$ between charged particles had 
to be taken into account. Referring to the valencies of monomer ions and counterions 
as $q_\alpha=\pm 1$, respectively, the following equation holds:
\begin{align}
\frac{V_{\rm Coul}(r_{\alpha\beta})}{k_{\rm B}T}&=
\frac{1}{k_{\rm B}T}\frac{q_\alpha q_\beta e^2}{\epsilon r_{\alpha\beta}}
\nonumber\\
&=\lambda_{\rm B}\frac{q_\alpha q_\beta}{r_{\alpha\beta}},
\end{align}
where $r_{\alpha\beta} = |{\bf r}_{\alpha}-{\bf r}_{\beta}|$, with ${\bf r}_{\alpha}$ 
and ${\bf r}_{\beta}$ denoting the position vectors of particles $\alpha$ and $\beta$, 
respectively. The Bjerrum length is defined as the length at which the electrostatic 
energy equals the thermal energy:
\begin{equation}
\lambda_{\rm B}=\frac{e^2}{\epsilon k_{\rm B}T},
\end{equation}
where $e$ is the unit charge and $\epsilon$ the permittivity of the solvent. For the 
case of water at room temperature one obtains $\lambda_{\rm B}=7.1$ {\AA}. In this 
work, we consider salt-free solutions only. Furthermore, the solvent was solely taken 
into account via its dielectric constant $\epsilon$. 
The Bjerrum length was fixed to 
$\lambda_{\rm B}=3.0\sigma_{\rm LJ}$.

The edge length $L$ of the simulation box was set having regard to the simulations' 
respective parameter combinations to suppress surface effects a priori. The center 
of mass of the system was fixed in the geometric middle of the box. The typical time 
step was $\Delta t=0.005\tau$, with 
$\tau=\sqrt{m\sigma_{\rm LJ}^2/\varepsilon_{\rm LJ}}$ being the associated time unit 
and $m$ the monomer mass. The counterions were taken to have the same mass and size 
as the charged monomers.

After a long equilibration time (typically $10^5$ time steps), different static 
quantities were calculated during simulation runs lasting about $5\times 10^5$ time 
steps. We carried out simulations for charging fractions $\alpha=0.10$ to $0.20$ 
and degrees of polymerization $N=50$ to $200$. In doing so, systematic predictions 
concerning the $\alpha$- and $N$-dependence of both all relevant conformational 
properties and the effective interaction became possible.

\subsection{
\label{sec:iso.t}
Theory
}

\begin{figure}[t]
\includegraphics[width=6cm,draft=false,clip]{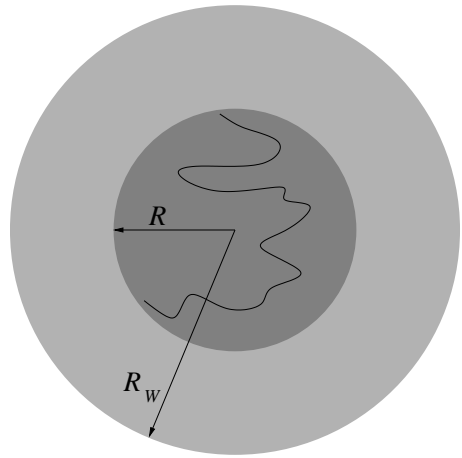}
\caption{\label{fig:iso1}Konieczny, Likos, L\"owen}
\end{figure}

In the theoretical investigations of the scaling properties of isolated PE chains, 
we use a mean-field, Flory-type approach, similar to that put forward in 
Refs.\ \onlinecite{klein.mm32}, \onlinecite{aj.prl} and \onlinecite{aj.jcp}. We 
consider a dilute, salt-free solution of density $\rho_{\rm chain}=N_{\rm chain}/V$,
where $N_{\rm chain}$ denotes the number of chains in the macroscopic volume $V$. 
For sake of simplicity, we take all ion species to be monovalent. Within the model, 
a single chain of total charge $Q$ is depicted as a spherical object of characteristic 
spatial extent $R$. The $N_{\rm c}$ counterions form an oppositely charged background, 
whereas we assume the corresponding density profile to be also spherically symmetric. 
The typical length scale $R_{\rm W}=\left(4\pi\rho_{\rm chain}/3\right)^{-1/3}$ of 
the counterion distribution is determined by $\rho_{\rm chain}$ alone. 
Fig.\ \ref{fig:iso1} shows a sketch of the situation, 
with the shaded domains visualizing 
the relevant length scales of our problem. 

In the following steps, we always restrict  our problem to the consideration of a 
single PE chain and its associated counterions. This simplification is convenient, 
since the $N_{\rm c}$ counterions of total charge $-Q$ assure electro-neutrality. 
Particular attention has to be paid to the Manning condensation of counterions on 
the semi-flexible chains.\cite{oosawa,dautzenberg,manning:jcp:69,deserno:macrom:00}
Let $b$ denote the average distance of two adjacent monomer ions along the chain 
backbone. Condensation takes place when the dimensionless fraction $\lambda_{\rm B}/b$ 
exceeds unity. The inequality $\lambda_{\rm B}/b<1$ holds for all our parameter 
combinations and according to Manning theory condensation effects do not play any 
role. It should be a reasonable approximation to neglect them completely in what 
follows. This expectation has been confirmed by our simulations.

Our main goal is to obtain theoretical predictions for the equilibrium expectation 
value of $R$. It is determined through minimization of a variational free energy, 
which reads as
\begin{equation}
\label{eq:f1}
\mathcal{F}^{(1)}(R;R_{\rm W})=U_{\rm H}^{(1)}+F_{\rm el}+F_{\rm int}+S_{\rm c}^{(1)}.
\end{equation}
Here, $U_{\rm H}^{(1)}$ is a mean-field electrostatic contribution, $F_{\rm el}$ 
and $F_{\rm int}$ represent the influences of conformational entropy and self 
avoidance of the monomers and $S_{\rm c}^{(1)}$ is the entropic contribution of 
the counterions. All terms will be described in more detail in what follows.

The electrostatic mean-field energy $U_{\rm H}^{(1)}$ is given by a Hartree-type 
expression of the general form
\begin{equation}
\label{eq:uh1.1}
U_{\rm H}^{(1)}=\frac{1}{2\epsilon}\iint {\rm d}^3r\; {\rm d}^3r'\; 
\frac{\varrho^{(1)}({\bf r})\varrho^{(1)}({\bf r'})}{\left|{\bf r}-{\bf r'}\right|},
\end{equation}
with the local charge density $\varrho^{(1)}({\bf r})$ to be defined below. On 
purely dimensional grounds, we expect a result that writes as
\begin{equation}
\label{eq:uh1.2}
\beta U_{\rm H}^{(1)} = \frac{N_{\rm c}^2\lambda_{\rm B}}{R}\cdot\vartheta^{(1)}
\left(\frac{R}{R_{\rm W}}\right),
\end{equation}
with $\beta=1/k_{\rm B}T$ denoting the inverse temperature. The shape of the 
dimensionless function $\vartheta^{(1)}(x)$ depends on the underlying charge 
distribution $\varrho^{(1)}({\bf r})$ alone.

The elastic contribution $F_{\rm el}$ to the free energy is entropic in nature, 
written as\cite{gennes}
\begin{equation}
\beta F_{\rm el} = \beta F(0) + \frac{3R^2}{2Na^2},
\end{equation}
with the equilibrium bond length $a$ and the unimportant constant $F(0)$. It 
stems from an ideal, i.e., Gaussian approximation of the conformational entropy 
of the chain. For the additional, non-electrostatic contribution, $F_{\rm int}$, 
of the PE chain, arising through self-avoidance, we employ the Flory-type
expression\cite{doi.edwards,gennes}
\begin{equation}
\label{eq:fint}
\beta F_{\rm int}(R) \sim v_0\frac{N^2}{2R^3}.
\end{equation}
Here, $v_0$ is the so-called excluded volume parameter, whose value has been 
discussed frequently.\cite{hariharan,pincus,odijk} The monomer volume turned 
out as a good approximation for the case of neutral polymer chains. Without 
Manning condensation, this guess remains valid for PE chains. Thus we chose 
$v_0\approx\sigma_{\rm LJ}^3$, where we identified the typical monomer length 
with $\sigma_{\rm LJ}$. Finally, the term $S_{\rm c}^{(1)}$ is an 
ideal entropic contribution of the form
\begin{align}
\label{eq:sc}
\beta S_{\rm c}^{(1)} = &\int{{\rm d}^3r\; \rho_{\rm c}(r)
\left[\ln \left( \rho_{\rm c}(r)\sigma_{\rm LJ}^3\right)-1 \right]}
+3N_{\rm c}\ln\left(\frac{\Lambda}{\sigma_{\rm LJ}}\right),
\end{align}
where $\Lambda$ is the thermal de-Broglie wavelength of the counterions. Since 
$S_{\rm c}^{(1)}$ is independent of the variational parameter $R$, it contributes
a trivial constant only and will we omitted in all further steps. It was solely 
included in Eq.\ \eqref{eq:f1} for the sake of completeness.

In order to carry out calculations following Eqs.\ \eqref{eq:f1} to \eqref{eq:fint}, 
assumptions concerning the number densities of all ionic particle species are required.
On this basis, the total charge density $\varrho^{(1)}({\bf r})$ can be computed. 
Due to the small charging fractions $\alpha$ considered here and according to 
polymer theory,\cite{cnl.habil, doi, doi.edwards} it is likely to expect a Gaussian 
shape of the local monomer density $\rho_{\rm m}({\bf r})$. For the counterions, an 
adequate approximation remains at first unknown. MD simulations can be helpful to 
estimate the local densities. Fig.\ \ref{fig:dens1} shows corresponding data 
confirming the above expectations concerning the monomer density profile and proving 
the absence of any condensation effects. The results regarding the counterion density 
are ambiguous. Admittedly, the concrete shape of the latter distribution is of minor 
importance for several reasons: in general, the local arrangement of particles 
is of secondary relevance when dealing with long-range forces. Since we investigate 
dilute solutions only, the counterion density remains low anyway. Moreover, without 
Manning condensation the effective charge of a PE chain equals the bare charge and 
we expect the monomer ions' contribution to electrostatic energy $U_{\rm H}^{(1)}$ 
to be dominant, while the influence of the counterions should be weak. Thus, our 
main guideline should be to preserve mathematical simplicity. In what follows, we 
will consider two varying sets of density profiles.

\subsubsection{
\label{sec:iso.ta}
Model A: Gaussian monomer density, homogeneous counterion background
}

Within the scope of a first approach, we postulate a Gaussian monomer density profile 
according to our MD simulation results
\begin{equation}
\label{eq:rhom1}
\rho_{\rm m}({\bf r})=\frac{N}{\pi^{3/2}R^{3}}e^{-(r/R)^2},
\end{equation} 
and furthermore assume the counterions to constitute a homogeneously charged 
background 
\begin{equation}
\rho_{\rm c}({\bf r})=\frac{N_{\rm c}\Theta(R_{\rm W}-r)}{V},
\end{equation}
where $\Theta(x)$ is the Heaviside step function. This leads to a total charge 
density $\varrho^{(1)}({\bf r})$, written as
\begin{align}
\label{eq:rho1}
\frac{\varrho^{(1)}({\bf r})}{Q} &= \left[ \frac{\rho_{\rm m}({\bf r})}{N}
-\frac{\rho_{\rm c}({\bf r})}{N_{\rm c}}\right]
\nonumber\\&
=\left[\frac{e^{-(r/R)^2}}{\pi^{3/2}R^3}-\frac{\Theta(R_{\rm W}-r)}{V}\right].
\end{align}

Therewith and following Eqs.\ \eqref{eq:uh1.1} and \eqref{eq:uh1.2}, we compute the 
dimensionless function $\vartheta^{(1)}(x)$. It reads as
\begin{align}
\vartheta^{(1)}(x)=&\frac{1}{\sqrt{2\pi}}+\frac{3x}{5}
-3x^2\left\{\frac{e^{-\left(1/x\right)^2}}{\sqrt{\pi}} 
-\left[ \frac{x}{2}-\frac{1}{x} \right]{\rm erf}\left(\frac{1}{x}\right)\right\}.
\end{align}

Summing up all contributions from Eq.\ \eqref{eq:f1} yields the total free 
energy ${\cal F}^{(1)}$ of the model system. The value $R$ determining the typical 
width of the spherical monomer cloud depicting the PE chain is then found by 
numerical minimization. It acquires an explicit density dependence through 
$R_{\rm W}$, as usual in charged systems.

\subsubsection{
\label{sec:iso.tb}
Model B: Gaussian monomer and counterion densities
}

We now introduce a different modeling of the considered system 
by replacing the 
homogeneous counterion background of Sec.\ \ref{sec:iso.ta} 
also with a Gaussian 
density profile, according to the equation
\begin{equation}
\rho_{\rm c}({\bf r})=\frac{N_{\rm c}}{\pi^{3/2}R_{\rm W}^{3}}e^{-(r/R_{\rm W})^2}.
\end{equation}
Regarding the monomer distribution, we still follow the assumptions of the previous 
section and define the local density $\rho_{\rm m}({\bf r})$ using 
Eq.\ \eqref{eq:rhom1}. For the overall charge density $\varrho^{(1)}({\bf r})$ we 
obtain analogous to Eq.\ \eqref{eq:rho1}:
\begin{equation}
\label{eq:rho1b}
\frac{\varrho^{(1)}({\bf r})}{Q}=\frac{1}{\pi^{3/2}} 
\left[\frac{e^{-(r/R)^2}}{R^3}-\frac{e^{-(r/R_{\rm W})^2}}{R_{\rm W}^3} \right].
\end{equation}

In anticipation of our considerations for two interacting PE chains [see 
Eq.\ (\ref{h2:eq}) in Sec.\ \ref{sec:int.t}], the following equation must hold:
\begin{equation}
U_{\rm H}^{(1)}(R;R_{\rm W})=\frac{1}{2} 
\lim_{D\rightarrow 0} U_{\rm H}^{(2)}(D;R;R_{\rm W}),
\end{equation}
where $U_{\rm H}^{(2)}(D;R;R_{\rm W})$ denotes the electrostatic mean-field 
contribution to the free energy of two PE chains with center-to-center separation 
$D$. Using the results of the Appendix, one finally gets
\begin{equation}
\vartheta^{(1)}(x)=\frac{1}{\sqrt{2\pi}}\left[1+x- 
\sqrt{\frac{8}{1+\left(\frac{1}{x}\right)^2}}\right].
\end{equation}

Compared to model A, all steps to come are completely analogous. Once more, 
minimization results obtained this way exhibit an explicit density dependence 
via $R_{\rm W}$.

\subsection{
\label{sec:iso.r}
Comparison to MD results
}

Based on the mean-field models presented above, predictions of the spatial extent 
of isolated PE chains become feasible. In order to verify this results, we have to 
test them against MD data and alternative theoretical approaches. In doing so, we 
first have to address a somewhat technical question. MD simulations yield 
expectation values for the radii of gyration $R_{\rm g}$ and the end-to-end 
distances $R_{\rm e}$, but neither of this quantities strictly corresponds to the 
variational parameter $R$ as obtained by minimization of the free energy 
${\cal F}^{(1)}$. Consequently, it remains unclear how a comparison can be carried 
out. Since we depicted the PE chains as spherical objects, one should expect $R$ 
to match half the end-to-end distance $R_{\rm e}/2$. Our results confirm such guess 
a posteriori, 
as will be seen shortly.

\begin{table}[t]
\caption{
\label{tab:iso.cmp}
Comparison of the chain radii of isolated PE chains obtained from both theory and 
MD simulation for different monomer numbers $N$ and charging fractions $\alpha$. 
The last column lists the number $N_{\rm c}^*$ of free counterions, necessary to 
calculate the effective chain interaction at nonoverlapping distances, see Eqs.\
\eqref{eq:v+1} and \eqref{eq:v+2}.
}
\begin{ruledtabular}
\begin{tabular}{ccccccc}
$N$ & $\alpha$ & $(R/\sigma_{\rm LJ})$\footnotemark[1] & 
$(R/\sigma_{\rm LJ})$\footnotemark[2] & 
$(R_{\rm g}/\sigma_{\rm LJ})$\footnotemark[3] & 
$(R_{\rm e}/\sigma_{\rm LJ})$\footnotemark[3] & 
$(N_{\rm c}^*)$\footnotemark[3]\\
\hline
50  & 0.10 & 10.0 & 10.2 & 5.7  & 14.9 & 4.8  \\
50  & 0.20 & 11.8 & 12.7 & 6.6  & 18.3 & 8.2  \\ 
100 & 0.10 & 17.2 & 17.5 & 10.4 & 27.1 & 9.4  \\
100 & 0.20 & 22.4 & 24.4 & 12.4 & 34.7 & 15.0 \\
150 & 0.10 & 24.4 & 24.9 & 13.9 & 41.9 & 13.5 \\
150 & 0.20 & 33.3 & 36.3 & 21.0 & 65.6 & 24.0 \\
200 & 0.10 & 31.7 & 32.5 & 23.7 & 65.9 & 19.4 \\
200 & 0.20 & 44.2 & 48.2 & 28.9 & 85.6 & 34.0
\end{tabular}
\end{ruledtabular}
\footnotetext[1]{Theory, model A.}
\footnotetext[2]{Theory, model B.}
\footnotetext[3]{MD simulation, $N_{\rm chain}=1$.}
\end{table}

\begin{figure}[b]
\begin{center}
\frame{\includegraphics[width=8cm,draft=false,clip]{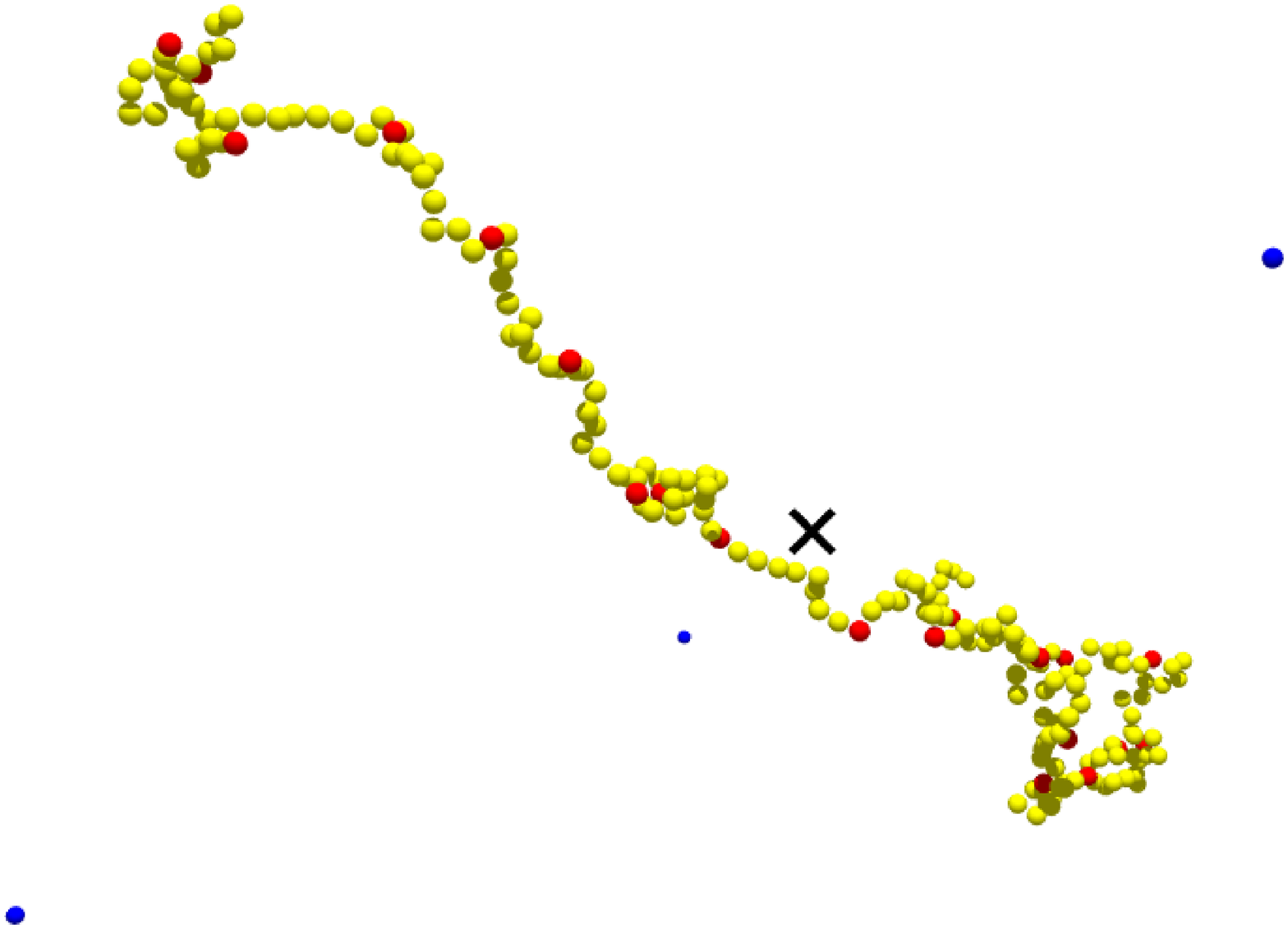}}
\frame{\includegraphics[width=8cm,draft=false,clip]{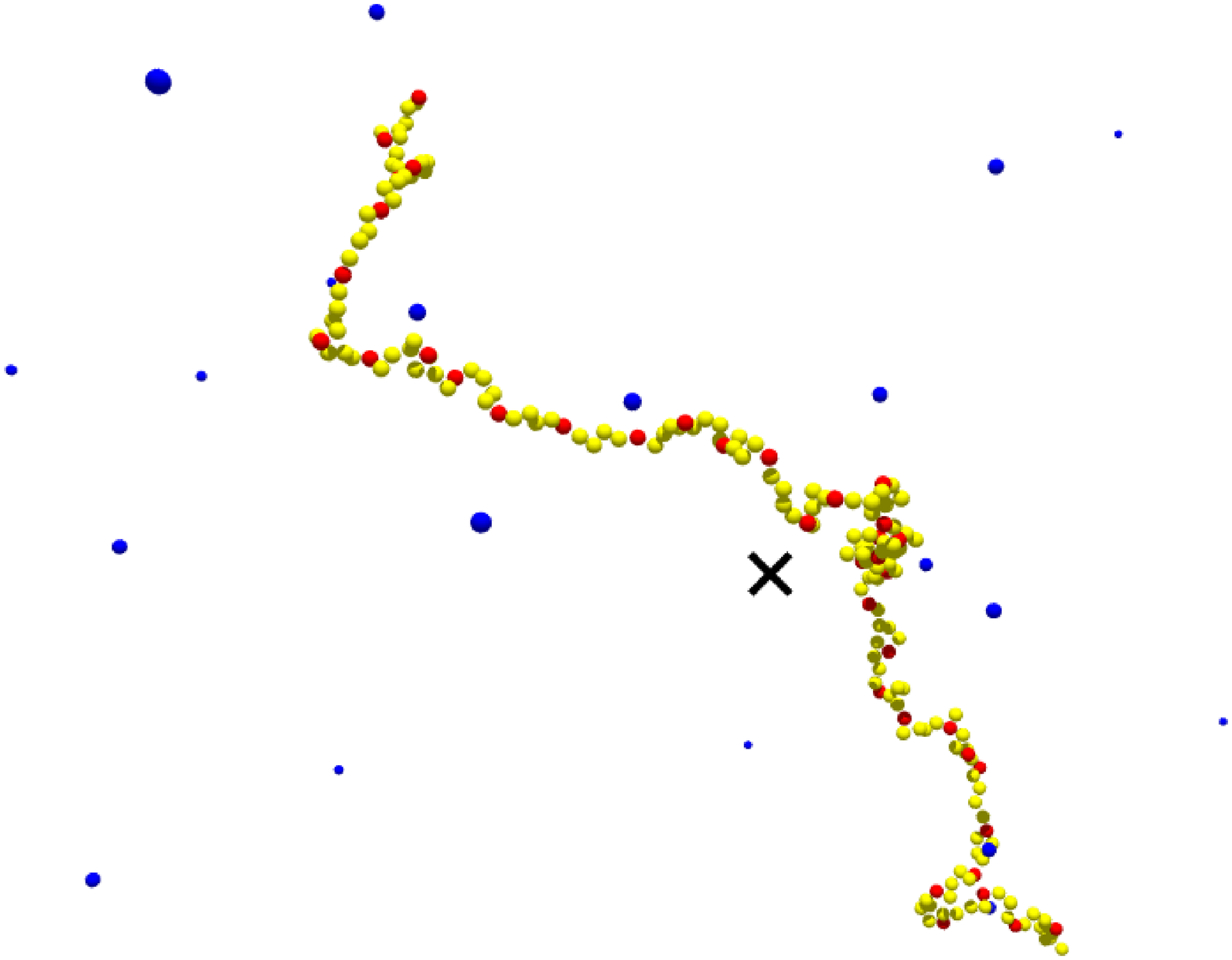}}
\end{center}
\vspace{-0.5cm}
\caption{\label{fig:iso.snapshot}Konieczny, Likos, L\"owen}
\end{figure}

Fig.\ \ref{fig:iso.snapshot} shows typical MD snapshots, that give rise to some 
first conclusions. A variation of the lateral chain extent becomes manifest, 
i.e., PE chains are more strongly stretched in the mid region than at their ends. 
This finding is in agreement with the results in Refs.\ \onlinecite{castelnovo} 
and \onlinecite{liao}. In general, the spatial configurations of PE chains are 
determined by two competing contributions to the free energy: the Coulomb repulsion 
and conformational entropy. For monomer ions in the middle of the chain sequence, 
electrostatic influences dominate and the described stretching is energetically 
favorable. But regarding the outermost parts of the chain backbone, the interaction 
of charged particles is of less importance, while the conformational entropy now 
becomes deciding. This leads to more coil-like arrangements of the affected 
monomers. In particular, short chains (monomer number $N\lesssim 50$) remain coiled 
even for increasing charging fractions $\alpha$. 

While simplified theoretical approaches predict a linear relationship $R\sim N$ 
between chain radius and monomer number,\cite{iff33} the effect described above 
requires some modifications of this scaling law. Considerations based on the concept 
of so-called electrostatic blobs yield weak logarithmic corrections, resulting 
in an expression of the general form\cite{castelnovo, migliorini, liao}
\begin{equation}
\label{eq:scale}
R\sim N(\ln N)^{1/3}.
\end{equation}

Fig.\ \ref{fig:iso.cmp1} compares MD data with chain radii obtained by both our 
mean-field theories and the scaling law pursuant to Eq.\ \eqref{eq:scale} for two 
different charging fractions. In the latter case, the unknown constant of 
proportionality was conveniently determined by means of a fit to the MD results. 
Thus, theoretical predictions and simulational findings are in good qualitative 
and quantitative agreement. This agreement particularly improves with increasing 
degree of polymerization $N$, because the basic assumption of continuous density 
profiles in place of an explicit consideration of discrete particles becomes more 
and more reasonable under these circumstances. Moreover, the presented results 
confirm our identification of the variational parameter $R$ with half the end-to-end 
distance $R_{\rm e}/2$. Here, we want to point out that the radius of gyration is 
noticeably smaller (see Tab.\ \ref{tab:iso.cmp}).

\begin{figure}[h]
\begin{center}
\includegraphics[width=8cm,draft=false,clip]{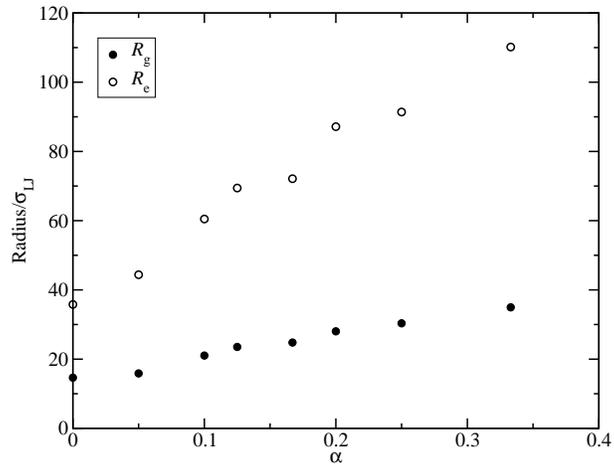}
\end{center}
\vspace{-0.5cm}
\caption{
\label{fig:iso.cmp2}Konieczny, Likos, L\"owen}
\end{figure}

In addition, Figs.\ \ref{fig:iso.cmp1} and \ref{fig:iso.cmp2} directly offer a 
possibility to read off information concerning the charge dependence of the 
spatial conformations of PE chains. According to this, in the event of very short 
chains, the chain radii are approximately unaffected by the charging fraction 
$\alpha$. Effects due to the Coulomb repulsion of equally charged beads do not 
clearly manifest themselves until the monomer number increases noticeably. Then, 
with increasing $\alpha$, a transition from coil-like to rod-like configurations 
takes place. In other words, for fixed $N$, the spatial extent of such chains is 
monotonously ascending with the charging fraction. Deviations from this behavior 
are likely to expect for highly charged chains only. Under these conditions, 
Manning condensation becomes more relevant and causes a collapse back to coil-like 
conformations.\cite{oosawa,dautzenberg,iff33,liu} However, for the range of parameters 
considered here, we do not have to take such processes into account, as our 
simulations unambiguously prove.

\section{
\label{sec:int}
Effective interactions between PE chains
}

\subsection{
\label{sec:int.t}
Theory
}

To derive a theoretical model for the effective interaction $V_{\rm eff}(D)$ of 
two PE chains, we separately consider the cases of small ($D\leq D_0$) and large 
($D>D_0$) distances of their centers of mass, respectively. The approach we apply 
is motivated by similar investigations for polymer and PE stars,\cite{aj.mm32,aj.jcp}
where an overlap condition for two stars of radius $R_{\rm star}$ determines the 
parameter $D_0=2R_{\rm star}$. But here, the linear chains we deal with are fractal 
objects of a certain flexibility. Thus, the notion of chain overlap in the previous 
sense becomes meaningless. The latter fact strongly affects our modeling. According 
to this, the requirement of continuity of the effective potential and the corresponding 
force when matching the partial solutions $V_{\rm eff}^-(D)$, valid for $D \leq D_0$, 
and $V_{\rm eff}^+(D)$, valid for $D > D_0$, defines the value of $D_0$. In particular,
$D_0$ varies depending on the set of parameters considered.

At first, we assume distances $D\leq D_0$. The effective interaction $V_{\rm eff}^-(D)$
results after taking a canonical trace over all but the chain center of mass degrees 
of freedom. Let ${\cal F}^{(2)}(z)$ be the Helmholtz free energy of a systems 
containing two PE chains at center-to-center separation $z$, therewith 
follows\cite{cnl.habil}
\begin{equation}
\label{eq:veff1}
V_{\rm eff}^-(D)={\cal F}^{(2)}(D)-{\cal F}^{(2)}(\infty).
\end{equation}

The term ${\cal F}^{(2)}(\infty)$ is manifestly independent of $D$ and contributes 
an additive constant only. On this account, it does not affect the effective 
forces, which are obviously obtained by deriving the effective potential with 
respect to $D$, i.e., using the equation 
$F_{\rm eff}(D)=-\partial V^-_{\rm eff}/\partial D$. Hence, we drop it in all 
steps to come.

\begin{figure}[t]
\begin{center}
\includegraphics[width=8cm,draft=false,clip]{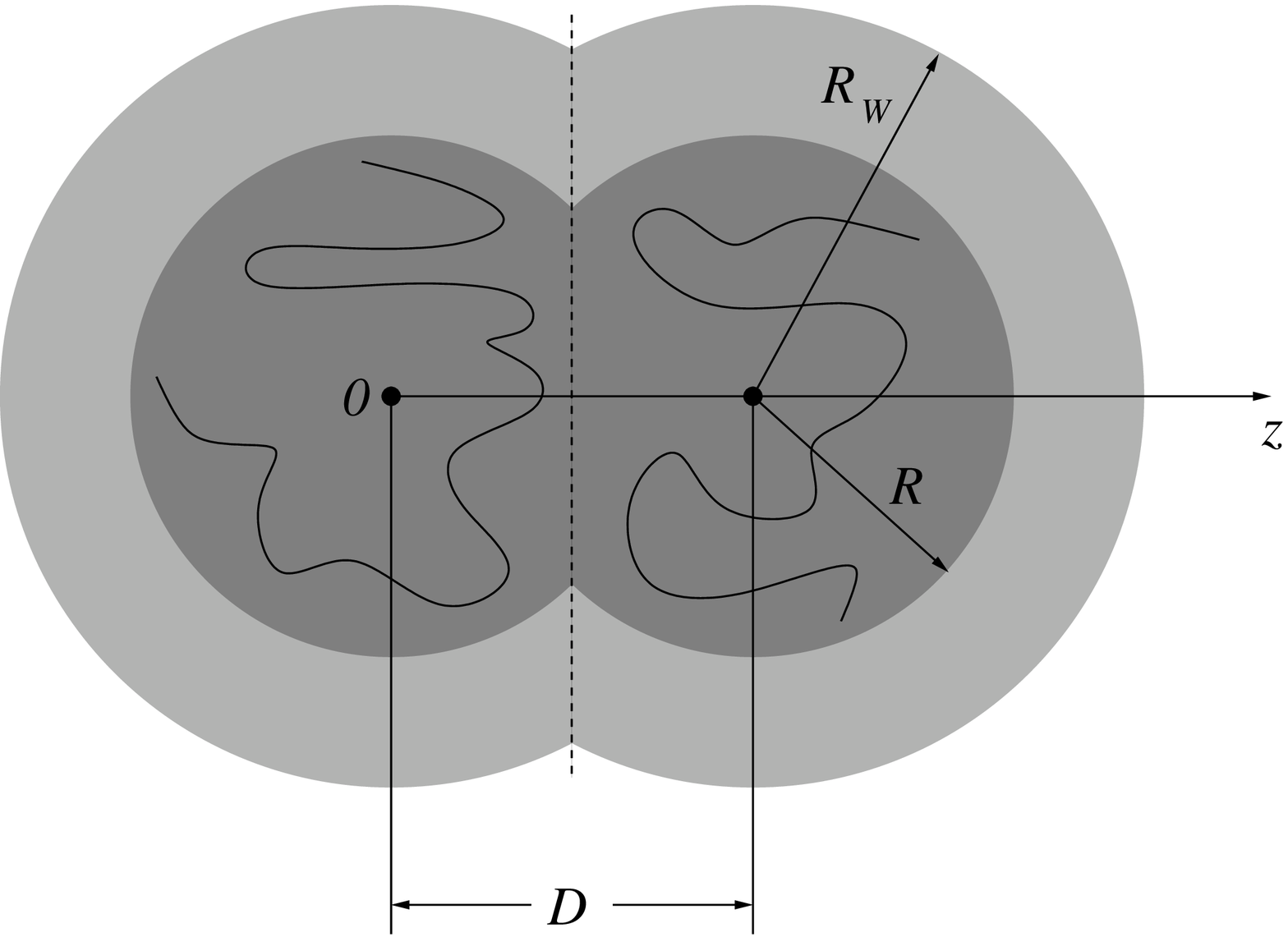}
\end{center}
\vspace{-0.5cm}
\caption{
\label{fig:int1}Konieczny, Likos, L\"owen}
\end{figure}

\begin{figure}[b]
\begin{center}
\includegraphics[width=8cm,draft=false,clip]{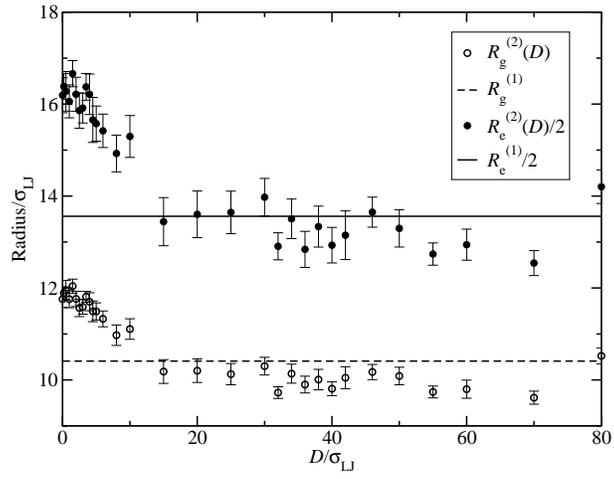}
\end{center}
\vspace{-0.5cm}
\caption{
\label{fig:int.radii}Konieczny, Likos, L\"owen}
\end{figure}

\begin{figure*}[t]
\begin{center}
\includegraphics[width=8cm,draft=false,clip]{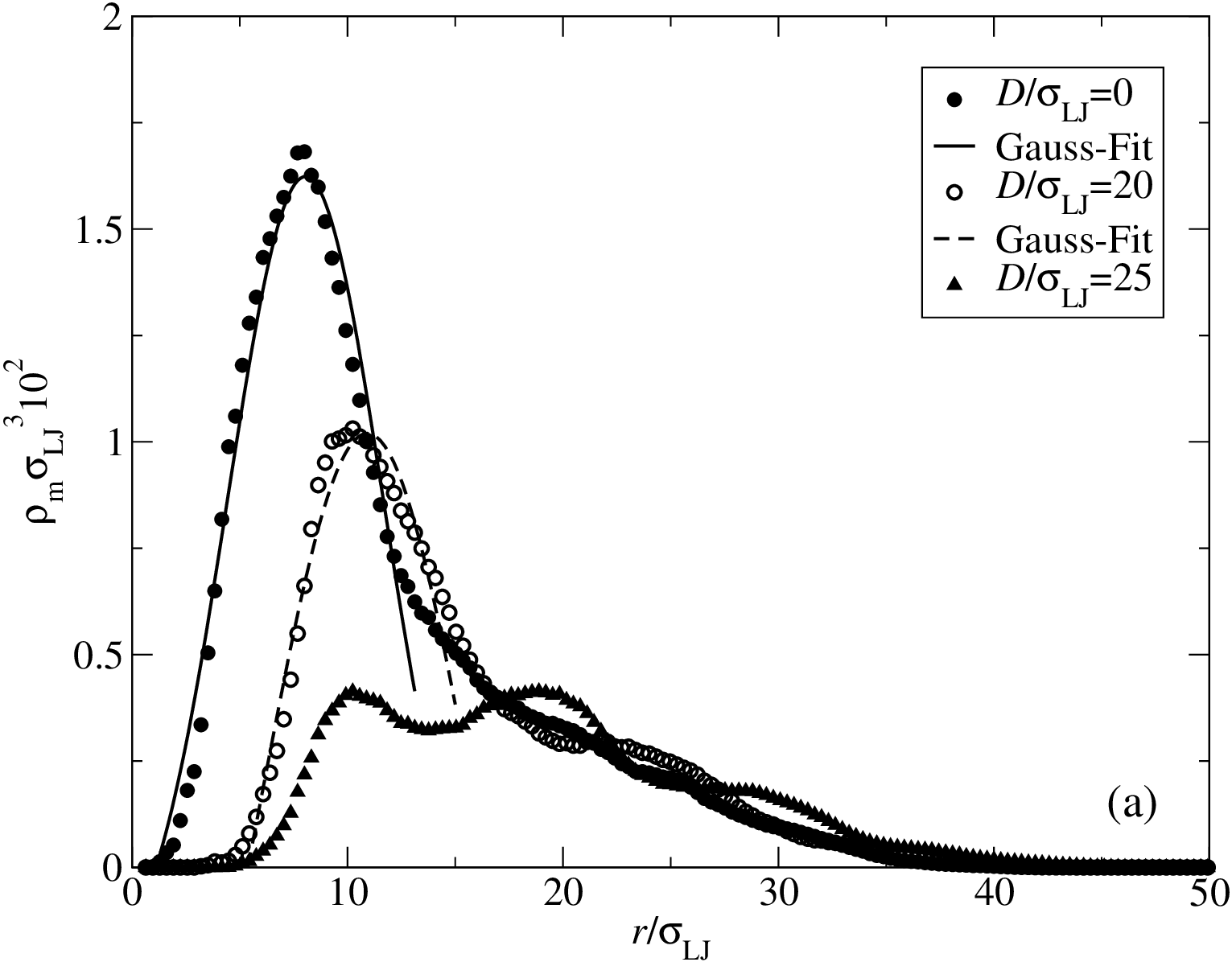}
\includegraphics[width=8cm,draft=false,clip]{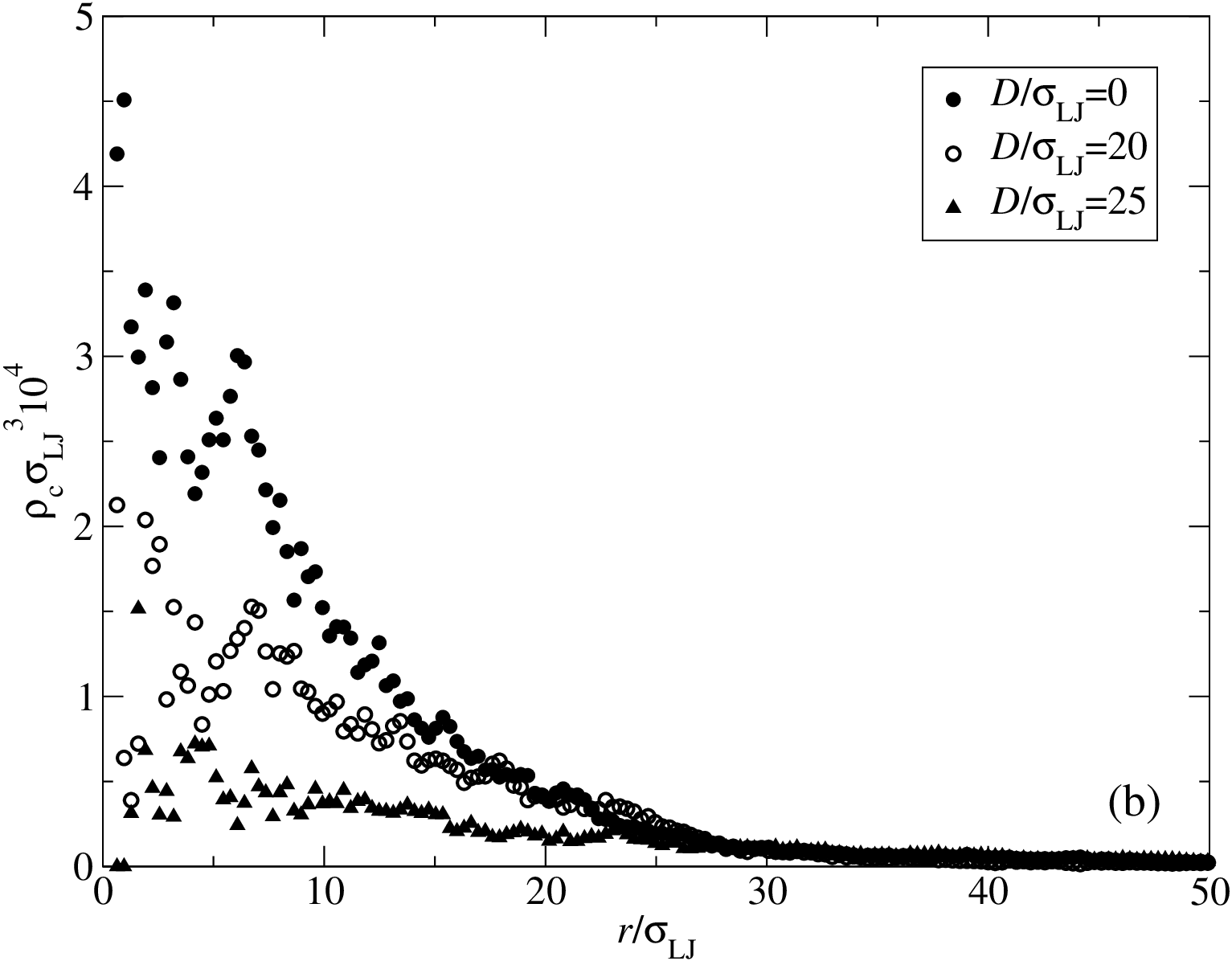}
\end{center}
\vspace{-0.5cm}
\caption{
\label{fig:int.charge}Konieczny, Likos, L\"owen}
\end{figure*}

To calculate the Helmholtz free energy ${\cal F}^{(2)}(D)$, we start from a 
mean-field approach analogous to Sec.\ \ref{sec:iso.tb} and model the system of
interacting PE chains by superposing two charge density profiles 
$\varrho^{(1)}({\bf r})$ as given via Eq.\ \eqref{eq:rho1b}. In other words, we 
assume them to maintain the same profiles as in the case of isolated chains and 
to completely interpenetrate, whereby both charge distributions are shifted with 
respect to each other by the vector ${\bf D}$ connecting the chains' centers of 
mass (see Fig.\ \ref{fig:int1}). Moreover, we expect the parameter $R$ to be 
independent of the center-to-center separation $D$. Again, the characteristic 
length scale $R_{\rm W}$ is determined by the density $\rho_{\rm chain}$ directly, 
i.e., by the equation $R_{\rm W}=\left(4\pi\rho_{\rm chain}/3\right)^{-1/3}$.  
As can be seen in Fig.\ \ref{fig:int.radii}, the radii of the chains remain essentially
unchanged even at full overlap. Moreover, in Fig.\ \ref{fig:int.charge} we show the
monomer density profiles, averaged over both chains. These are measured during the
MD runs as a function of the distance from their common center of mass, for various
separations $D$ between the centers of mass of the two chains. It can bee seen that
for distances $D$ as small as $10\sigma_{\rm LJ}$, the individual profiles can still
be described by Gaussian functions. Consequently, the overall charge density 
$\varrho^{(2)}({\bf r};{\bf D})$ reads as
\begin{equation}
\label{eq:rho2}
\varrho^{(2)}({\bf r};{\bf D})=\varrho^{(1)}({\bf r})+\varrho^{(1)}({\bf r}-{\bf D}).
\end{equation}

By now, we are interested in the effective interaction as a function of the 
center-to-center separation $D$ exclusively, whereas conformational properties 
should not be considered. Thus, contrary to Sec.\ \ref{sec:iso.t}, we omit any 
minimization of the free energy. Instead, we use the width $R$ of the monomer 
density profiles as additional fit parameter. As we will see, optimal agreement 
between theoretical predictions and MD data is achieved when $R$ approximately 
matches half the end-to-end distance $R_{\rm e}/2$ of isolated PE chains (see 
Tab.\ \ref{tab:int.cmp}).

In what follows, we completely neglect the conformational entropy $F_{\rm el}$ and 
excluded volume interaction $F_{\rm int}$, which are only weakly dependent on 
the inter-chain distance $D$. Hence, they yield constant contributions only and 
do not influence the effective potential $V_{\rm eff}(D)$ significantly. Our 
problem immediately reduces to the computation of electrostatic term 
$U_{\rm H}^{(2)}(D)$ and counterion entropy $S_{\rm c}^{(2)}(D)$, i.e., we have
\begin{equation}
{\cal F}^{(2)}(D;R;R_{\rm W})=U_{\rm H}^{(2)}(D)+S_{\rm c}^{(2)}(D).
\end{equation}

Taking Eqs.\ \eqref{eq:rho1b} and \eqref{eq:rho2} into account, we are able to 
conveniently rewrite Eq.\ \eqref{eq:uh1.1}. Dropping all irrelevant $D$-independent 
self energy terms, we obtain
\begin{equation}
U_{\rm H}^{(2)}(D)=
\frac{1}{\epsilon} \iint{\vol{^3r}\;\vol{^3r'}\;
\frac{\varrho^{(1)}({\bf r})\varrho^{(1)}({\bf r'}-{\bf D})}{\left|{\bf r}-{\bf r}'\right|}}.
\label{h2:eq}
\end{equation}
On purely dimensional grounds, the result of the above integration must read as 
(cf.\ Sec.\ \ref{sec:iso.t})
\begin{equation}
\beta U_{\rm H}^{(2)}(D)=
\frac{N_c^2\lambda_{\rm B}}{D}\cdot\vartheta^{(2)}\left(\frac{R}{D},\frac{R_{\rm W}}{D}\right).
\end{equation}



Once more, the specific shape of the dimensionless function $\vartheta^{(2)}(x)$ 
depends on the underlying charge densities $\varrho^{(1)}({\bf r})$ and 
$\varrho^{(2)}({\bf r})$ alone. A detailed derivation in case of Gaussian 
density profiles for each chain and for the associated counterions (Model B) according 
to Eq.\ \eqref{eq:rho1b} of Sec.\ \ref{sec:iso.tb} is presented in the Appendix. The 
corresponding result reads as
\begin{align}
\label{eq:int.uh}
\vartheta^{(2)}\left(\frac{R}{D},\frac{R_{\rm W}}{D}\right)=
&h\left(\frac{R^2}{2D^2}\right)+h\left(\frac{R_{\rm W}^2}{2D^2}\right)
-2h\left(\frac{R^2+R_{\rm W}^2}{4D^2}\right),
\end{align}
where we introduced the abbreviation
\begin{equation}
h(x)=\frac{2}{\pi}\int_{0}^{\infty}{{\rm d}t\;\frac{\sin(t)}{t}e^{-t^2x}}.
\end{equation}


Since an analytical computation of both $h(x)$ and the counterion entropy 
$S_{\rm c}(D)$ is not feasible, we need to make use of numerical 
methods,\cite{num.recip} starting from Eqs.\ \eqref{eq:sc}, \eqref{eq:rho1b} 
and \eqref{eq:rho2}. By means of this, the Helmholtz free energy 
${\cal F}_2(D;R;R_{\rm W})$ and the effective potential $V_{\rm eff}^-(D)$ 
are known in principle. They exhibit a typical explicit density dependence 
via $R_{\rm W}$. The corresponding forces must inherit such property, what 
has to be regarded during any comparison to MD data.

\begin{figure}[b]
\begin{center}
\includegraphics[width=8cm,draft=false,clip]{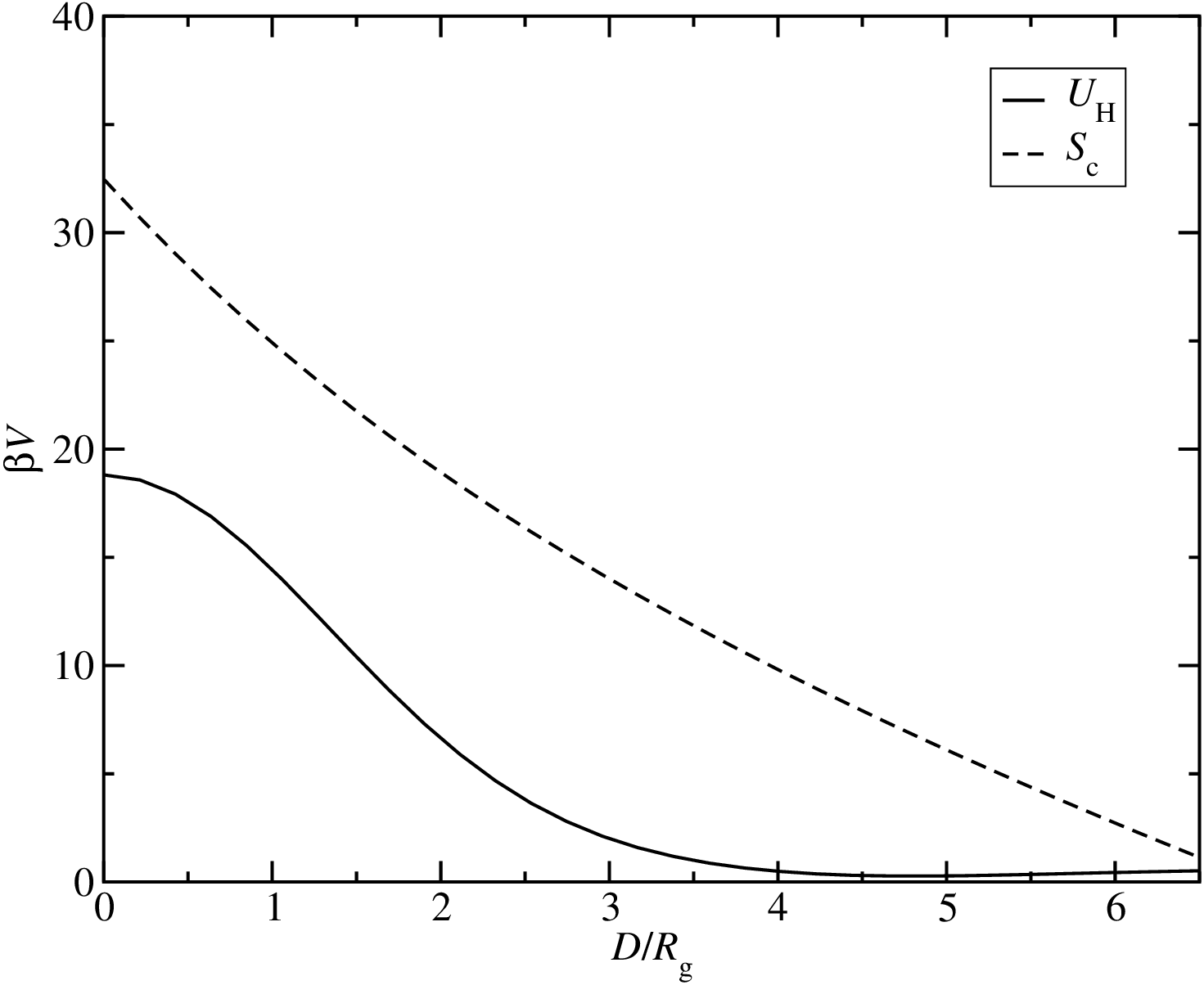}
\end{center}
\vspace{-0.5cm}
\caption{
\label{fig:int.uhvssc}Konieczny, Likos, L\"owen}
\end{figure}

Fig.\ \ref{fig:int.uhvssc} quantitatively compares the parts $U_{\rm H}^{(2)}(D)$ 
and $S_{\rm c}^{(2)}(D)$ for an exemplary chosen parameter combination. 
It can be seen that their variations with $D$, which determines the effective force
through the $D$-derivative of these contributions, are comparable, i.e., neither 
of them dominates the physics of our system over the other. This is contrary to 
comparable results for PE stars,\cite{aj.jcp} where the electrostatic energy 
constitutes the major influence for small arm numbers $f$, while it is of minor 
importance for the opposite case of big $f$. For the latter situation, the PE stars 
nearly act as if the were neutral, due to the strong adsorption of counterions in
their interior,\cite{aj.prl,aj.jcp} a mechanism absent in the case of PE chains
at hand.

Now, the case of inter-chain distances $D>D_0$ has to be considered. We assume 
the chains to be spheres of average radius $R_{\rm g}$, whose respective net 
charges $Q^*=N^*_{\rm c}e$ are reduced compared to the bare charges $Q=N_{\rm c}e$ 
due to counterions located within this spheres. Each chain is surrounded by a 
cloud of the remaining $N_{\rm c}^*$ free counterions, which cause an additional 
electrostatic screening of the interaction. Again, 
$R_{\rm W}=\left(4\pi\rho_{\rm chain}/3\right)^{-1/3}$ denotes the clouds' 
characteristic spatial extent.

Adopting a Debye-H{\"u}ckel approach\cite{lowen:review} for nonoverlapping
polyelectrolytes, we postulate a Yukawa-type expression for the effective 
interaction potential there, i.e.,
\begin{equation}
\label{eq:v+1}
\beta V_{\rm eff}^+(D)=N_{\rm c}^2\lambda_{\rm B}\frac{e^{-D/\lambda_{\rm D}}}{D}.
\end{equation}
Here, the screening length $\lambda_{\rm D}$ is mainly determined by the radii 
$R_{\rm g}$ of the spheres and the number $N_{\rm c}^*$ of free counterions. We 
obtain\cite{cnl.habil}
\begin{equation}
\label{eq:v+2}
\lambda_{\rm D}=
\left[\frac{R_{\rm W}^3-R_{\rm g}^3}{3N^*_{\rm c}\lambda_{\rm B}}\right]^{1/2}.
\end{equation}



In what follows, we fix both $R_{\rm g}$ and $N_{\rm c}^*$ according to MD results 
(see Tab.\ \ref{tab:iso.cmp}). Thereby, the latter is determined by calculating the
time average of the number of counterions within an imaginary sphere around a single chain's 
center of mass whose radius equals the instantaneous value of the corresponding radius 
of gyration. In doing so, we know the mean net charge $Q^*$ at the same time. There are no 
further fit parameters, i.e., the Yukawa tail $V_{\rm eff}^+(D)$ of the effective potential 
is completely specified by Eqs.\ \eqref{eq:v+1} and \eqref{eq:v+2}. It exhibits explicit 
and implicit density dependences via the quantities $R_{\rm W}$ and $N^*_{\rm c}$, 
respectively.


In order to determine the interval boundary $D_0$, we simply introduce a 
continuity constraint for the effective force 
$F_{\rm eff}(D)=-\partial V_{\rm eff}(D)/\partial D$. In mathematical formulation, 
the corresponding condition reads as
\begin{equation}
\left.\frac{\partial\left[V_{\rm eff}^-(D)-V_{\rm eff}^+(D)\right]}{\partial D}\right|_{D_0}=0.
\end{equation}
The final step in deriving the total potential $V_{\rm eff}(D)$ is to match both 
parts of the function at $D=D_0$. This is feasible by constantly shifting 
$V_{\rm eff}^-(D)$ only, since $V_{\rm eff}(D)$ must vanish for increasing chain 
separations $D\rightarrow\infty$. Thus, the following definition is convenient:
\begin{equation}
V_{\rm eff}(D)=
\begin{cases}
V_{\rm eff}^-(D)+V_{\rm eff}^+(D_0)-V_{\rm eff}^-(D_0),&D\leq D_0\\
V_{\rm eff}^+(D),&D>D_0
\end{cases}.
\end{equation}

Fig.\ \ref{fig:int.veff} shows corresponding results for several parameter 
combinations. The effective interaction is purely repulsive in nature, ultra-soft 
and even bounded for vanishing distances $D\rightarrow 0$. There is a strong 
charge dependence of the potential, i.e., changing the charging fraction from 
$\alpha=0.10$ to $\alpha=0.20$ roughly yields a factor two. To emphasize the latter 
fact, Fig.\ \ref{fig:int.veff}(c) additionally includes an effective potential 
$V_{\rm eff}^{(0)}$ for neutral polymer chains, which was computed using the 
equation\cite{louis:pre}
\begin{equation}
\beta V_{\rm eff}^{(0)}(D)
=1.87\cdot\exp\left[-\left(\frac{D}{1.13\cdot R_{\rm g}}\right)^2\right].
\end{equation}
Compared to charged systems, the neutral polymer effective interaction is about 
two orders of magnitude weaker. Within the scope of our considerations, we did 
not investigate the explicit and implicit density dependences of the effective 
interaction potential in more detail.

\subsection{
\label{sec:int.r}
Comparison to MD results
}

Now, our theoretical model has to be tested against MD results. A straightforward 
comparison is not feasible, since (effective) interaction potentials are not 
accessible directly using standard (MD) simulation techniques. By means of the 
expressions
\begin{align}
\label{eq:feff.m}
{\bf F}^{\rm (m)}_\alpha({\bf r}_{1i_0},{\bf r}_{2i_0})
&=-\nabla_{{\bf r}_{\alpha i_0}}V_{\rm eff}({\bf r}_{1 i_0},{\bf r}_{2 i_0})\nonumber\\
&=\left<{\bf f}_{\alpha i_0}\right>_D\nonumber\\
&={\bf F}^{\rm (m)}_\alpha(D)
\end{align}
and
\begin{align}
\label{eq:feff.c}
{\bf F}^{\rm (c)}_\alpha({\bf R}_1,{\bf R}_2)
&=-\nabla_{{\bf R}_\alpha}V_{\rm eff}({\bf R}_1,{\bf R}_2)\nonumber\\
&=\left<\sum_{i=1}^{N}{{\bf f}_{\alpha i}}\right>_D\nonumber\\
&={\bf F}^{\rm (c)}_\alpha(D),
\end{align}
respectively, we can measure the mean forces acting on given monomers $i_0$ or 
the centers of mass. Here, let ${\bf f}_{\alpha i}$ be the instantaneous force 
that monomer $i$ of chain $\alpha$ experiences and ${\bf r}_{\alpha i}$ the 
corresponding spatial position. In addition, the vectors ${\bf R}_\alpha$
denote the chains' respective centers of mass. In both equations above, 
$\left<\ldots\right>_D$ denotes an averaging with fixed chain distance
$D=\left|{\bf r}_{1i_0}-{\bf r}_{2i_0}\right|$ or $D=\left|{\bf R}_1-{\bf R}_2\right|$.
For symmetry reasons, the following relation must hold:
\begin{equation}
{\bf F}^{\rm (m,c)}_1(D)=-{\bf F}^{\rm (m,c)}_2(D)
\end{equation}

While the validity of Eq.\ \eqref{eq:feff.m} is manifest, Eq.\ \eqref{eq:feff.c} 
needs a mathematical proof, described in detail in Ref.\ \onlinecite{goetze}. 
In all further steps, we consider the magnitude of the force. It is connected to 
the potential via
\begin{align}
F_{\rm eff}(D)&=\left|{\bf F}^{\rm (m,c)}_{1,2}(D)\right|\nonumber\\
&=-\frac{\partial V_{\rm eff}}{\partial D}.
\end{align}
According to this, predictions for the effective forces become computable 
starting from theoretical results for the corresponding potential. This allows 
the desired comparison to MD data in principle.

Choosing the centers of mass as effective coordinates is arbitrary; one
could have picked, e.g., the central monomers as representatives of the
whole chain. The concrete choice of effective coordinates does affect the 
effective force, as physically expected. However, once the separation 
between the centers of mass or the central monomers exceeds the typical chain
size, it is reasonable to expect that it should be largely irrelevant
which degrees of freedom are kept fixed, since the chains appear at
these scales as diffuse objects. The latter assumption has been confirmed 
by our MD simulations. Following this argument, we limit ourselves to the 
consideration of the centers of mass as effective coordinates according to 
Eq.\ \eqref{eq:feff.c}.

The mean-field model presented in Sec.\ \ref{sec:int.t} predicts ultra-soft and 
even bounded effective potentials $V_{\rm eff}(D)$ (Fig.\ \ref{fig:int.veff}). 
Thus, the corresponding forces reach their respective maxima at inter-chain 
distances $D/R_{\rm g}>0$, which is in good quantitative agreement with MD results. 
Fig.\ \ref{fig:int.feff} shows a concluding comparison of theoretical results and 
MD data for various parameter combinations. For the lower charge fraction, 
$\alpha = 0.10$, we find good agreement between theory and simulation, although 
we observe qualitative deviations for very small inter-chain distances. There, 
the theoretically predicted forces are in part stronger and, in particular, 
do not drop to zero for $D/R_{\rm g}\rightarrow 0$, as one would expect on
symmetry grounds. This discrepancy is due to the use of Gaussian density profiles 
(cf.\ Sec.\ \ref{sec:int.t}), since such approximation becomes unreasonable for 
vanishing chain separations (see Fig.\ \ref{fig:int.charge}). The quality of the 
theory decreases for the larger value of charge fraction, $\alpha = 0.20$, since 
the chains in this case become more stretched and the underlying theoretical assumption 
of a Gaussian monomer profile starts losing its validity. Here, modeling the chains as 
rods would probably yield better results, although these rods are not completely stiff, 
i.e., pronounced lateral fluctuations are still present. For all plots, values of the 
fit parameter $R$ (in case $D\leq D_0$) are explicitly given. They roughly equal the 
half of the end-to-end distances $R_{\rm e}$ of the chains, as measured within 
simulations (cf.\ Tab.\ \ref{tab:iso.cmp}). This again allows an interpretation 
of $R$ as spatial extent of the PE chains and establishes the consistency of the 
approach for the effective interaction with that for the isolated chains.

The interval boundary $D_0$ is identifiable by means of a slight cusp in the force 
vs.\ distance curves. The latter is not physically reasonable, but an artifact 
arising as consequence of the matching of both branches. Without the introduction 
of at least one additional fit parameter, such effect is not avoidable. Our results 
exhibit a distinct charge dependence of $D_0$. While $D_0/R_{\rm g}\approx 2$ holds 
for $\alpha=0.10$ and independent of the monomer number $N$, the position of the 
matching moves towards bigger distances $D_0/R_{\rm g}\approx 4$ for $\alpha=0.20$ 
and arbitrary $N$.

\begin{table}[t]
\caption{
\label{tab:int.cmp}
Comparison of the fit parameter $R$ to the chain radii of isolated PE chains as 
obtained by MD simulation for different monomer numbers $N$ and charging fractions 
$\alpha$.
}
\begin{ruledtabular}
\begin{tabular}{ccccc}
$N$ & $\alpha$ & $(R/\sigma_{\rm LJ})$\footnotemark[1] &  
$(R_{\rm g}/\sigma_{\rm LJ})$\footnotemark[2] & 
$(R_{\rm e}/2\sigma_{\rm LJ})$\footnotemark[2] \\
\hline
50  & 0.10 & 6.1  & 5.7  & 7.5  \\
50  & 0.20 & 9.0  & 6.6  & 9.2  \\ 
100 & 0.10 & 12.7 & 10.4 & 13.6 \\
100 & 0.20 & 20.5 & 12.4 & 17.4 \\
150 & 0.10 & 19.4 & 13.9 & 21.0 \\
150 & 0.20 & 31.4 & 21.0 & 32.8 \\
200 & 0.10 & 24.7 & 23.7 & 33.0 \\
200 & 0.20 & 44.0 & 28.9 & 42.8 
\end{tabular}
\end{ruledtabular}
\footnotetext[1]{Theory.}
\footnotetext[2]{MD simulation, $N_{\rm chain}=1$.}
\end{table}

\section{
\label{sec:con}
Summary and conclusions
}

We have presented a theoretical approach in describing the conformations
and sizes of polyelectrolyte chains. The theory is based on the
assumption of Gaussian monomer- and counterion-profiles around the
chain's center of mass and employs a variational free energy that
includes electrostatic, excluded-volume and entropic contributions.
By direct comparison with simulation results, we have shown that
the theory is capable of predicting the typical size of PE chains
that are weakly charged and its dependence on the charge fraction
and the degree of polymerization. Thereafter, the theory has been
extended to describe two interacting polyelectrolytes and the 
effective interaction potential between their centers of mass has
been derived and compared successfully with computer simulation
results. In doing so, we have made a first step in describing
polyelectrolytes as soft colloids, in analogy with recently developed
approaches for neutral polymers.\cite{louis:prl, louis:pre, louis:jcp} 
The effective potentials obtained are ultrasoft and bounded but
nevertheless much more repulsive than those obtained for polymer
chains. The physical reasons lie both in the electrostatic repulsion
between the charges carried on the chains and on the entropically
caused osmotic pressure of the counterions that are trapped
within the interior of the chains.

The effective potential includes an explicit density dependence
arising from the redistribution of counterions inside and outside
the chains upon a change of the overall concentration. In principle,
this potential can be employed in order to study the structural 
characteristics of concentrated solutions of polyelectrolytes,
i.e., the correlations between the centers of mass of the chains.
Moreover, it can be used for the calculation of thermodynamic
properties of concentrated solutions, such as free energies and
pressure and of phase transitions, such as the phase separation
investigated recently by computer simulations.\cite{thanassis:prl:03}
Indeed, whereas it has been already shown that the approach of
coarse-graining successfully predicts phase separation in 
polymer blends,\cite{archer:jpcm:02} the question of whether the
same is true for polyelectrolytes has not been examined to date.
We plan to return to this problem in the future.

\begin{acknowledgments}
We thank Arben Jusufi, Ren{\'e} Messina, Norman Hoffmann and Christian Mayer 
for helpful discussions. Financial support from the DFG (SFB TR6) is
acknowledged.
\end{acknowledgments}

\appendix*

\section{
\label{sec:app.uh}
Proof of Eq.\ (\ref{eq:int.uh})
}

In Secs.\ \ref{sec:iso} and \ref{sec:int}, we dropped technical details of the 
calculation of the Helmholtz free energies. Here, we want to carry out the 
derivation of the mean-field contribution $U_{\rm H}^{(2)}(D)$ for interacting 
PE chains (cf.\ Sec.\ \ref{sec:int.t}) explicitly. We start from the 
equation
\begin{equation}
\label{eq:a1}
U_{\rm H}^{(2)}(D)=\frac{1}{2\epsilon}\iint{\vol{^3r}\;\vol{^3r'}\;\frac{\varrho^{(2)}({\bf r};{\bf D})\varrho^{(2)}({\bf r}';{\bf D})}
{\left|{\bf r}-{\bf r}'\right|}}.
\end{equation}
Using \eqref{eq:rho2} in \eqref{eq:a1} and neglecting all irrelevant self-energy 
terms, which are independent of $D$, we get
\begin{equation}
\label{eq:a2}
U_{\rm H}^{(2)}(D)=\frac{1}{\epsilon}\iint{\vol{^3r}\;\vol{^3r'}\;\frac{\varrho^{(1)}({\bf r})\varrho^{(1)}({\bf r'}-{\bf D})}{\left|{\bf r}-{\bf r}'\right|}}.
\end{equation}
Let $*$ denote a convolution of two functions. With the definition 
$v(r)=1/(\epsilon r)$, we rewrite \eqref{eq:a2} obtaining
\begin{align}
\label{eq:a3}
U_{\rm H}^{(2)}(D)&=\int{\vol{^3r'}\;\varrho^{(1)}({\bf r'}-{\bf D})\int{\vol{^3r}\;v(\left|{\bf r}-{\bf r}'\right|)\varrho^{(1)}({\bf r})}}\nonumber\\
&=\int{\vol{^3r'}\;\varrho^{(1)}({\bf r'}-{\bf D})\left[\varrho^{(1)}*v\right]({\bf r})}\nonumber\\
&=\left[\varrho^{(1)}*\left[\varrho^{(1)}*v\right]\right]({\bf D}).
\end{align}
Now, we pass into Fourier space and thereby introduce the notation $\tilde{f}$ 
for the associated Fourier transform of a function $f$. First of all, this yields 
the relation
\begin{equation}
\label{eq:a4}
\left[\varrho^{(1)}*v\right]({\bf r})=\frac{1}{\left(2\pi\right)^3}\int{\vol{^3k}\;\tilde{\varrho}^{(1)}({\bf k})\tilde{v}({\bf k})
e^{{\rm i}{\bf k}\cdot{\bf r}}}.
\end{equation}
Taking (\ref{eq:a4}) into account, (\ref{eq:a3}) reads as
\begin{align}
U_{\rm H}^{(2)}(D)&=\left[\varrho^{(1)}*\left[\varrho^{(1)}*v\right]\right]({\bf D})\nonumber\\
&=\frac{1}{\left(2\pi\right)^3}\int{\vol{^3k}\;\tilde{\varrho}^{(1)}({\bf k})\widetilde{\left[\varrho^{(1)}*v\right]}({\bf k})e^{{\rm i}{\bf k}\cdot{\bf D}}}
\nonumber\\
\label{eq:a5}
&=\frac{1}{\left(2\pi\right)^3}\int{\vol{^3k}\;\left[\tilde{\varrho}^{(1)}({\bf k})\right]^2\tilde{v}({\bf k})e^{{\rm i}{\bf k}\cdot{\bf D}}}.
\end{align}
With $\varrho^{(1)}({\bf r})/Q=\rho_{\rm m}({\bf r})/N-\rho_{\rm c}({\bf r})/N_{\rm c}$
and regarding the linearity of the Fourier transform, Eq.\ \eqref{eq:a5} writes as
\begin{widetext}
\begin{align}
U_{\rm H}^{(2)}(D)
=&\frac{Q^2}{\left(2\pi\right)^3}\left\{\frac{1}{N^2}\int{\vol{^3k}\;\tilde{\rho}^2_{\rm m}({\bf k})\tilde{v}({\bf k})e^{{\rm i}{\bf k}\cdot{\bf D}}}
+\frac{1}{N_{\rm c}^2}\int{\vol{^3k}\;\tilde{\rho}^2_{\rm c}({\bf k})\tilde{v}({\bf k})e^{{\rm i}{\bf k}\cdot{\bf D}}}\right.\nonumber\\
\label{eq:a6}
&\left.-\frac{2}{NN_{\rm c}}\int{\vol{^3k}\;\tilde{\rho}_{\rm m}({\bf k})\tilde{\rho}_{\rm c}({\bf k})\tilde{v}({\bf k})e^{{\rm i}{\bf k}\cdot{\bf D}}}\right\}.
\end{align}
\end{widetext}
Up to now, we did not assume any specific properties of the density profiles or the 
integration kernel $v$. All considerations are valid in a very general fashion. In 
our special case, the mentioned functions are radially symmetric. Due to this fact, 
we obtain
\begin{widetext}
\begin{align}
U_{\rm H}^{(2)}(D)=&\frac{Q^2}{2\pi^2D}\left\{\frac{1}{N^2}\int_{0}^{\infty}{\vol{k}\;k\sin(kD)\tilde{\rho}_{\rm m}^2(k)\tilde{v}(k)}
+\frac{1}{N^2_{\rm c}}\int_{0}^{\infty}{\vol{k}\;k\sin(kD)\tilde{\rho}_{\rm c}^2(k)\tilde{v}(k)}\right.\nonumber\\
&\left.-\frac{2}{NN_{\rm c}}\int_{0}^{\infty}{\vol{k}\;k\sin(kD)\tilde{\rho}_{\rm m}(k)\tilde{\rho}_{\rm c}(k)}\tilde{v}(k)\right\}\nonumber\\
\label{eq:a7}
=&\frac{Q^2}{2\pi^2D}\left\{I_1(D)+I_2(D)-2I_3(D)\right\}
\end{align}
\end{widetext}
Starting from Eqs.\ \eqref{eq:rho1b}, \eqref{eq:rho2} and using the above definition 
of $v$, we find
\begin{align}
\label{eq:a8}
\frac{\tilde{\rho}_{\rm m}(k)}{N}&=e^{-k^2R^2/4},\\
\label{eq:a9}
\frac{\tilde{\rho}_{\rm c}(k)}{N_{\rm c}}&=e^{-k^2 R_{\rm W}^2/4},\\
\label{eq:a10}
\tilde{v}(k)&=\frac{4\pi}{\epsilon k^2}.
\end{align}
Based on \eqref{eq:a7} and including Eqs.\ \eqref{eq:a8} to \eqref{eq:a10}, we 
firstly have as an intermediate result
\begin{align}
I_1(D)&=\frac{4\pi}{\epsilon}\int_{0}^{\infty}{\vol{k}}\;\frac{\sin(kD)}{k}e^{-k^2R^2/2}\nonumber\\
&=\frac{4\pi}{\epsilon}\int_{0}^{\infty}{\vol{t}}\;\frac{\sin(t)}{t}e^{-t^2(R^2/2D^2)}\nonumber\\
\label{eq:a11}
&=\frac{2\pi^2}{\epsilon}\cdot h\left(\frac{R^2}{2D^2}\right).
\end{align}
In complete analogy, we achieve the relations
\begin{align}
\label{eq:a12}
I_2(D)&=\frac{2\pi^2}{\epsilon}\cdot h\left(\frac{ R_{\rm W}^2}{2D^2}\right)
\end{align}
and moreover
\begin{align}
\label{eq:a13}
I_3(D)&=\frac{2\pi^2}{\epsilon}\cdot h\left(\frac{R^2+ R_{\rm W}^2}{4D^2}\right).
\end{align}
For simplicity reasons, we thereby introduced the abbreviated notation
\begin{equation}
\label{eq:a14}
h\left(x\right)=\frac{2}{\pi}\int_{0}^{\infty}{\vol{t}}\;\frac{\sin(t)}{t}e^{-t^2x}.
\end{equation}
Furthermore, we define based on the previous Eqs.\ \eqref{eq:a11} to \eqref{eq:a13}:
\begin{widetext}
\begin{align}
\label{eq:a15}
\vartheta^{(2)}\left(\frac{R}{D},\frac{ R_{\rm W}}{D}\right)&=\frac{\epsilon}{2\pi^2}\left\{I_1(D)+I_2(D)-2I_3(D)\right\}\nonumber\\
&=\left\{h\left(\frac{R^2}{2D^2}\right)+h\left(\frac{ R_{\rm W}^2}{2D^2}\right)
-2h\left(\frac{R^2+ R_{\rm W}^2}{4D^2}\right)\right\}.
\end{align}
\end{widetext}
Substituting \eqref{eq:a15} in \eqref{eq:a7} finally yields the result
\begin{align}
\label{eq:a16}
\beta U_{\rm H}^{(2)}(D)&=\frac{\beta Q^2}{\epsilon D}\cdot\vartheta^{(2)}\left(\frac{R}{D},\frac{ R_{\rm W}}{D}\right)\nonumber\\
&=\frac{N_c^2\lambda_{\rm B}}{D}\cdot\vartheta^{(2)}\left(\frac{R}{D},\frac{ R_{\rm W}}{D}\right).
\end{align}

\clearpage

\begin{center}
FIGURE CAPTIONS
\end{center}

FIG.\ 1: A sketch showing a PE chain of typical spatial extent $R$ and the 
surrounding counterion sphere of radius $R_{\rm W}$.

FIG.\ 2: Local number densities of (a) the monomers and (b) of the counterions 
for an isolated PE chain with $N=200$, $\alpha=0.10$ and $R_{\rm W}/\sigma_{\rm LJ}=124.1$, 
as obtained by MD simulations. Here $r$ denotes the distance from the center 
of mass of the chain.

FIG.\ 3: MD snapshots of isolated PE chains with degree of polymerization 
$N=200$ and two different charging fractions $\alpha=0.10$ (left) and 
$\alpha=0.20$ (right). Bright gray balls are neutral monomers, and the dark 
spheres along the chain indicate the charged monomers. The counterions are 
the small, dark gray spheres around the chain. In addition, the black crosses 
mark the centers of mass of the instantaneous configurations.

FIG.\ 4: Spatial extent of isolated PE chains as a function of the degree of 
polymerization $N$ for charging fractions: (a) $\alpha=0.10$ and (b) $\alpha=0.20$. 
The lines show theoretical results while the symbols denote data obtained by 
our MD simulations.

FIG.\ 5: Chain radii of isolated PE chains as a function of the charging fraction 
$\alpha$ for fixed monomer number $N=200$ and $R_{\rm W}/\sigma_{\rm LJ}=124.1$, 
as obtained by our MD simulations.

FIG.\ 6: A sketch of two overlapping PE chains, for a more detailed explanation 
of the theoretical modeling, see the text.

FIG.\ 7: Simulation results for the radii of interacting PE chains, depending 
on their center of mass separation $D$. The results shown here refer to parameters 
$N=100$ and $\alpha=0.10$. The corresponding radii of isolated chains are shown 
as horizontal lines for comparison.

FIG.\ 8: Density profiles (a) for the monomers and (b) for the counterions of 
interacting PE chains with $N=200$, $\alpha=0.10$, $R_{\rm W}/\sigma_{\rm LJ}=157.6$ 
and varying center of mass separation $D$, as obtained via MD simulation. 

FIG.\ 9: Comparison of the terms $U_{\rm H}^{(2)}(D)$ and $S_{\rm c}^{(2)}(D)$ 
that contribute to the Helmholtz free energy as functions of the chain separation 
$D$, where we exemplarily chose $N=200$, $\alpha=0.10$ and  $R_{\rm W}/\sigma_{\rm LJ}=157.6$.

FIG.\ 10: Theoretically predicted effective potential $V_{\rm eff}(D)$ as a function 
of the interchain separation $D$ for (a) $N=100$, (b) $N=150$, and (c) $N=200$. 
The corresponding charging fractions are $\alpha=0.10$ and $\alpha=0.20$, as 
indicated in the legends. For all cases, the used values of the fit parameter 
$R$ are specified in the legend boxes. In panel (c) we additionally 
show as the dash-dotted line the result valid for neutral polymers.\cite{louis:pre}



FIG.\ 11: Simulation snapshots of interacting PE chains with degree of polymerization 
$N=200$ and charging fraction $\alpha=0.20$. Bright gray balls are neutral
monomers, and the dark spheres along the chains indicate the charged monomers. 
The counterions are the small, dark gray spheres around the chains. The distances
of centers of mass, which are marked by the black crosses, are 
$D=2\sigma_{\rm LJ}$ (left) and $D=40\sigma_{\rm LJ}$ (right), respectively.
In general, the mean chain directions are not perpendicular to the vector
${\bf D}$ connecting the centers of mass.


FIG.\ 12: The effective force $F_{\rm eff}(D)$ on the center of mass of a polyelectrolyte 
as a function of the interchain separation $D$, and for charging fractions 
$\alpha=0.10$ [(a), (b), (c)] and $\alpha=0.20$ [(d), (e), (f)], where the respective 
degrees of polymerization are: $N=100$ [(a), (d)], $N=150$ [(b), (e)],
and $N = 200$ [(c), (f)]. Solid lines are theoretical predictions, with the fit 
parameters $R$ specified in the legend boxes. Data points denote MD results for the
chains' centers of mass as reference points.

\clearpage

\clearpage

\begin{figure}[b]
\begin{center}
\includegraphics[width=8cm, draft=false]{eps/dens.m_Nchain.1_N.200_alpha.10.eps}
\includegraphics[width=8cm, draft=false]{eps/dens.ci_Nchain.1_N.200_alpha.10.eps}
\end{center}
\vspace{-0.5cm}
\caption{\label{fig:dens1}Konieczny, Likos, L\"owen}
\end{figure}

\clearpage

\clearpage

\begin{figure*}[h]
\begin{center}
\includegraphics[width=8cm,draft=false,clip]{eps/radii_N.var_boxdim.var_alpha.10.eps}
\includegraphics[width=8cm,draft=false,clip]{eps/radii_N.var_boxdim.var_alpha.20.eps}
\end{center}
\vspace{-0.5cm}
\caption{
\label{fig:iso.cmp1}Konieczny, Likos, L\"owen}
\end{figure*}

\clearpage

\clearpage

\clearpage

\clearpage

\clearpage

\clearpage

\begin{figure*}[t]
\begin{center}
\includegraphics[width=5.4cm,draft=false,clip]{eps/Veff_N.100_alpha.var.eps}
\includegraphics[width=5.4cm,draft=false,clip]{eps/Veff_N.150_alpha.var.eps}
\includegraphics[width=5.4cm,draft=false,clip]{eps/Veff_N.200_alpha.var.eps}
\end{center}
\vspace{-0.5cm}
\caption{
\label{fig:int.veff}Konieczny, Likos, L\"owen}
\end{figure*}

\clearpage

\begin{figure}[b]
\begin{center}
\frame{\includegraphics[width=8cm,draft=false,clip]{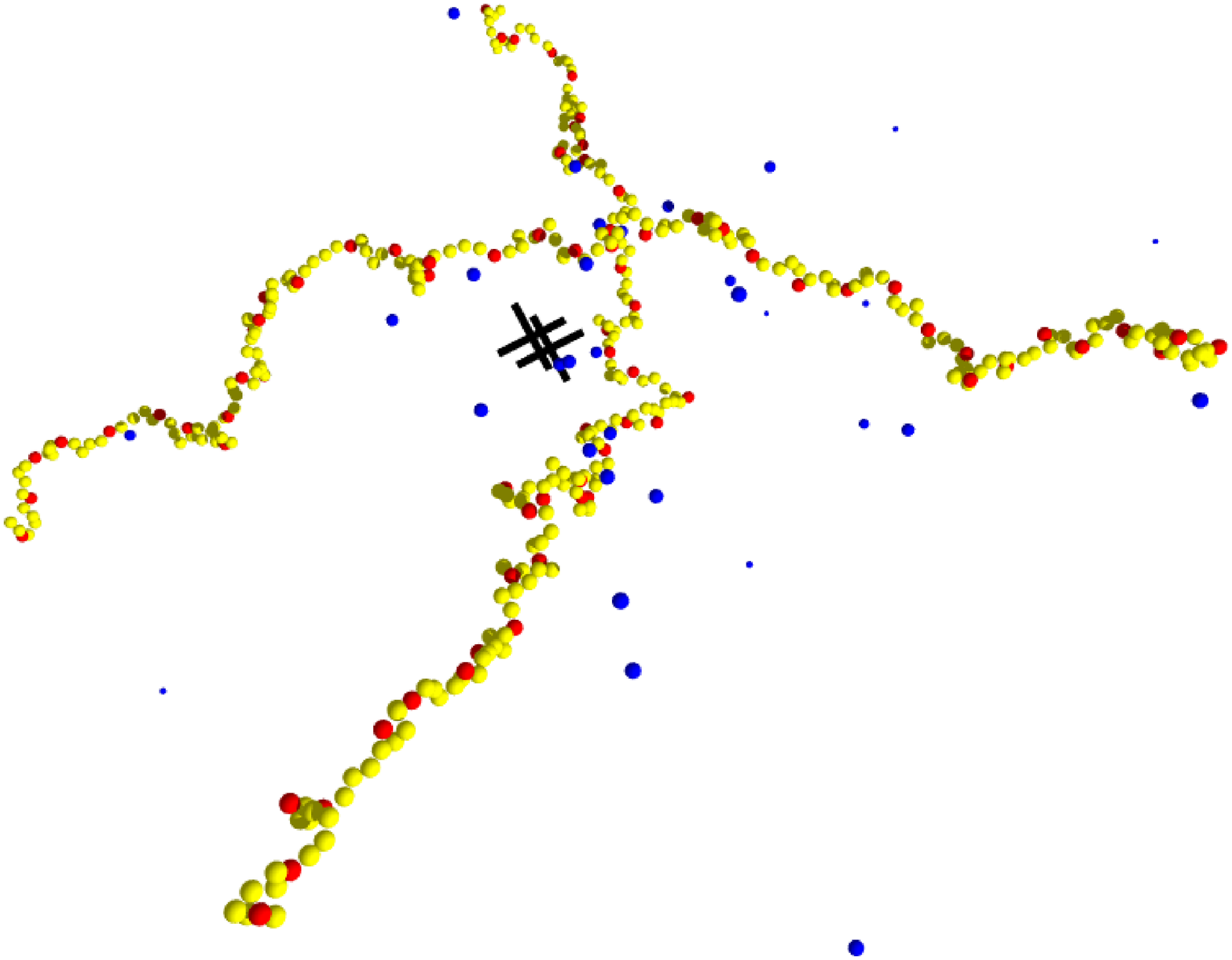}}
\frame{\includegraphics[width=8cm,draft=false,clip]{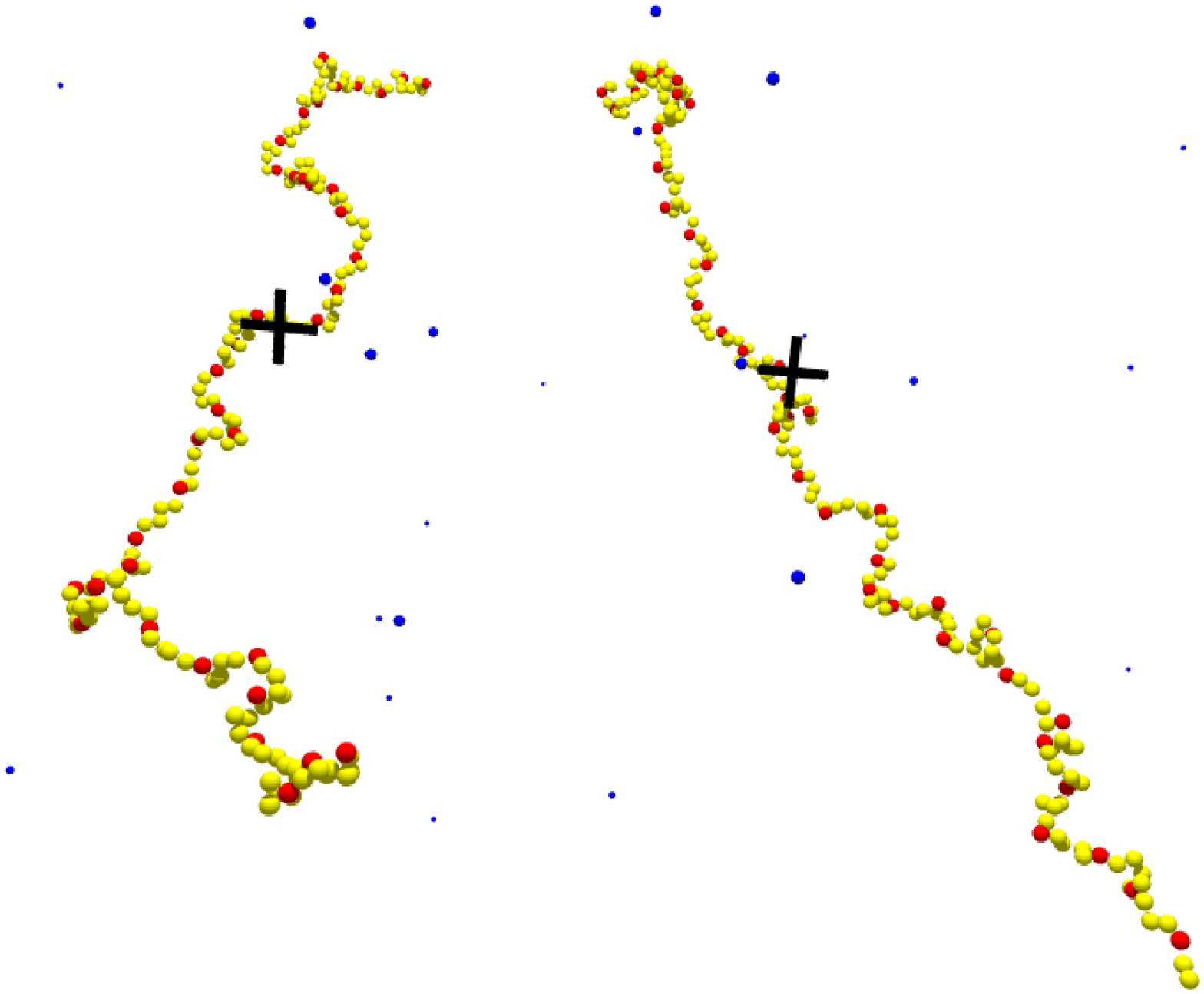}}
\end{center}
\vspace{-0.5cm}
\caption{\label{fig:int.snapshot}Konieczny, Likos, L\"owen}
\end{figure}

\clearpage

\begin{figure*}[t]
\begin{center}
\includegraphics[width=5.4cm,draft=false,clip]{eps/Feff_N.100_alpha.10.eps}
\includegraphics[width=5.4cm,draft=false,clip]{eps/Feff_N.150_alpha.10.eps}
\includegraphics[width=5.4cm,draft=false,clip]{eps/Feff_N.200_alpha.10.eps}
\end{center}
\begin{center}
\includegraphics[width=5.4cm,draft=false,clip]{eps/Feff_N.100_alpha.20.eps}
\includegraphics[width=5.4cm,draft=false,clip]{eps/Feff_N.150_alpha.20.eps}
\includegraphics[width=5.4cm,draft=false,clip]{eps/Feff_N.200_alpha.20.eps}
\end{center}
\vspace{-0.5cm}
\caption{
\label{fig:int.feff}Konieczny, Likos, L\"owen}
\end{figure*}

\end{document}